\documentclass[lettersize,journal]{IEEEtran}
\usepackage{amsmath,amsfonts}
\usepackage{algorithmic}
\usepackage{array}
\usepackage[caption=false,font=normalsize,labelfont=sf,textfont=sf]{subfig}
\usepackage{textcomp}
\usepackage{stfloats}
\usepackage{url}
\usepackage{verbatim}
\usepackage{graphicx}
\usepackage{bbm}
\usepackage{cancel}
\usepackage{float}
\usepackage{highlight}
\usepackage{booktabs} 
\usepackage{orcidlink}
\usepackage{amssymb}
\usepackage{nicematrix}
\usepackage{siunitx}
\usepackage{flushend}

\def\BibTeX{{\rm B\kern-.05em{\sc i\kern-.025em b}\kern-.08em
    T\kern-.1667em\lower.7ex\hbox{E}\kern-.125emX}}
\usepackage{balance}

\begin{document}

\bstctlcite{IEEEexample:BSTcontrol}

\title{Constrained Optimization of Charged Particle Tracking with Multi-Agent Reinforcement Learning}
\author{Tobias~Kortus\thanks{Tobias Kortus, Ralf Keidel and Nicolas R. Gauger are with the Scientific Computing Group, University of Kaiserslautern-Landau (RPTU), Kaiserslautern, Germany. (e-mails: ralf.keidel@rptu.de; tobias.kortus@rpu.de, nicolas.gauger@scicomp.uni-kl.de)},~Ralf~Keidel, Nicolas R. Gauger, Jan Kieseler\thanks{Jan Kieseler is with the Institute of Experimental Particle Physics (ETP), Karlsruhe Institute of Technology (KIT), Karlsruhe, Germany (jan.kieseler@kit.edu)}, \\ Bergen pCT Collaboration}
%\thanks{Manuscript created October, 2020; This work was developed by the IEEE Publication Technology Department. This work is distributed under the \LaTeX \ Project Public License (LPPL) ( http://www.latex-project.org/ ) version 1.3. A copy of the LPPL, version 1.3, is included in the base \LaTeX \ documentation of all distributions of \LaTeX \ released 2003/12/01 or later. The opinions expressed here are entirely that of the author. No warranty is expressed or implied. User assumes all risk.}}

%\markboth{Journal of \LaTeX\ Class Files,~Vol.~18, No.~9, September~2020}%
%{How to Use the IEEEtran \LaTeX \ Templates}

\markboth{PRE-PRINT (2025-JAN-09) / THIS PAPER IS CURRENTLY UNDER REVIEW}%
{}

\maketitle

\begin{abstract}
Reinforcement learning demonstrated immense success in modelling complex physics-driven systems, providing end-to-end trainable solutions by interacting with a simulated or real environment, maximizing a scalar reward signal. In this work, we propose, building upon previous work, a multi-agent reinforcement learning approach with assignment constraints for reconstructing particle tracks in pixelated particle detectors. Our approach optimizes collaboratively a parametrized policy, functioning as a heuristic to a multidimensional assignment problem, by jointly minimizing the total amount of particle scattering over the reconstructed tracks in a readout frame. To satisfy constraints, guaranteeing a unique assignment of particle hits, we propose a safety layer solving a linear assignment problem for every joint action. Further, to enforce cost margins, increasing the distance of the local policies predictions to the decision boundaries of the optimizer mappings, we recommend the use of an additional component in the blackbox gradient estimation, forcing the policy to solutions with lower total assignment costs. We empirically show on simulated data, generated for a particle detector developed for proton imaging, the effectiveness of our approach, compared to multiple single- and multi-agent baselines. We further demonstrate the effectiveness of constraints with cost margins for both optimization and generalization, introduced by wider regions with high reconstruction performance as well as reduced predictive instabilities. Our results form the basis for further developments in RL-based tracking, offering both enhanced performance with constrained policies and greater flexibility in optimizing tracking algorithms through the option for individual and team rewards.
\end{abstract}


\section{Introduction}

\IEEEPARstart{R}{einforcement} learning (RL) and multi-agent reinforcement learning (MARL) are promising paradigms for constructing and optimizing autonomous agents that can compete in a wide variety of complex sequential decision problems such as games~\cite{Mnih2013, Silver2018}, robotics~\cite{Gu2017, Andrychowicz2020} or autonomous driving~\cite{Kendall2019} by discovering complex interaction mechanisms in the underlying environment. Coupled with the tremendous success in the aforementioned fields, RL has recently demonstrated great potential in optimizing and controlling physics processes~\cite{Degrave2022, Kain2020, Vage2022, Kortus2023}, by maximizing a scalar reward signal using trial and error~\cite{Sutton2018, Littman1994}. Especially for problems of combinatorial nature, RL demonstrated to be able to learn generalizable policies that are even able to outperform supervised learning approaches, despite the lack of ground truth information~\cite{Joshi2021}. \emph{Kortus et al.} \cite{Kortus2023} and \emph{V{\aa}ge} \cite{Vage2022} have shown for charged particle tracking used in high-energy physics reconstruction, the potential of deep reinforcement learning for optimizing over discrete assignment operations, aiming to construct discrete sets of particle tracks over subsequent layers under the influence of particle interaction mechanisms. Extending previous work, we further investigate the concept of RL-based charged particle tracking as a combinatorial optimization problem. We therefore propose a collaborative MARL approach with assignment constraints, iteratively optimizing a joint policy of multiple track follower. We represent the stepwise agent constraints as a centralized safety layer, ensuring unique hit assignment across all agents, both during training and inference, by solving a linear sum assignment problem (LSAP) projecting the unsafe local agent policies to a global safe policy. All source code together with  hyperparameters, data, and models are available on GitHub\footnote{\url{https://github.com/SIVERT-pCT/marl-tracking}} and Zenodo\footnote{\url{https://doi.org/10.5281/zenodo.7426388}}. Our main contributions and findings in this paper summarize as follows:

\begin{itemize}
    \item Building upon previous work in~\cite{Kortus2023}, we propose multiple multi-agent extensions of RL-based particle tracking, using decentralized agents with optional safety layer, satisfying assignment constraints, trained in a centralized manner using centralized critic architectures.
    \item Increasing the cost margins between predictions and decision boundaries efficiently, we extend the blackbox differentiation technique by~\cite{Vlastelica2020} by an additional simple gradient component, resulting in significantly improved training and generalization abilities.
    \item We demonstrate excellent empirical performance of our method, compared to a conventional track follower~\cite{Pettersen2020} as well as single-agent~\cite{Kortus2023} and multi-agent baselines.
    \item Finally, we validate the benefit of the architecture and adapted gradient through the safety layer by examining reconstruction performance, reward surfaces~\cite{Sullivan2022}, prediction instabilities~\cite{Mahdi2016}, and policy entropy.
\end{itemize}

\begin{figure*}
    \centering
    \includegraphics[width=14cm]{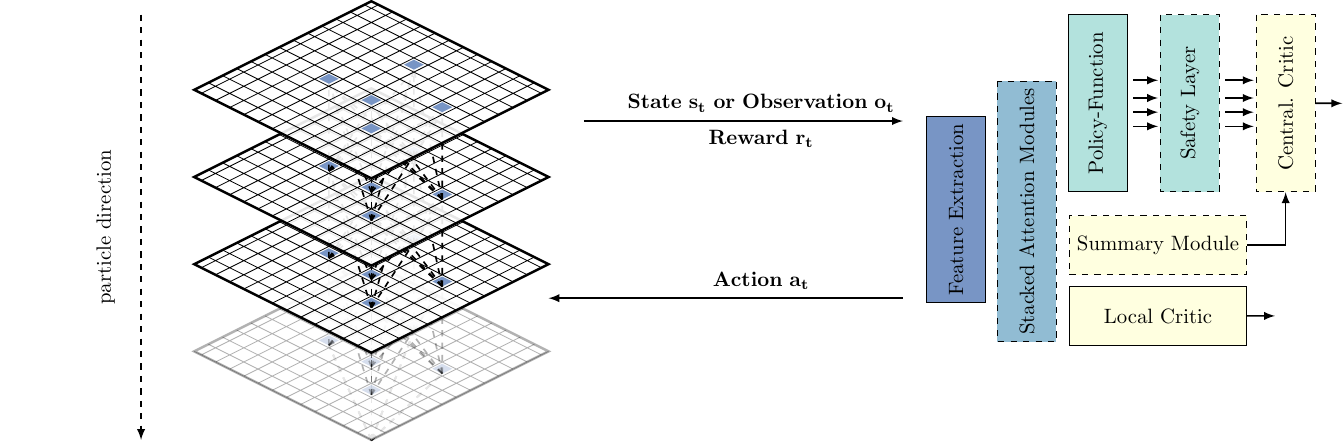}
    \caption{General description of charged particle tracking framework for single- or multi-agent reinforcement learning. The agent (right) learns by iterated interaction with the environment, represented as a directed acyclic graph (left), reconstruction policies that maximize the obtained rewards. Agent components marked with dashed lines are optional and are only used for some agent configurations.}
    \label{fig:architecture}
\end{figure*}

\section{Theory and Background}

Throughout this work, we focus on particle data generated by the digital tracking calorimeter (DTC) prototype developed by the \emph{Bergen pCT Collaboration} \cite{Alme2020, Aehle2023b} for proton computed tomography. In the following section, we describe both the detector and the basic particle interaction mechanisms expected at relevant particle energies of $\mathcal{O}$(\SI{230}{MeV}). \\

\paragraph{Bergen pCT detector prototype}  The Bergen pCT DTC is a multi-layer pixelated tracking calorimeter, consisting in total of two tracking layers and 41 detector-absorber sandwich calorimeter layers. It uses multiple strips of ALPIDE pixel sensors \cite{Mager2016, AglieriRinella2017} with additional \SI{3.5}{mm} aluminum absorbers in each calorimeter layer, for measuring and reconstructing particles stopping in the detector. Further details and a fine-grained decomposition of the detector material is described in~\cite{Alme2020}. While the exact composition of the detector is not essential for our work, we want to point out the different material budgets of the tracking and calorimeter layer. Both components are used in combination for accurate estimation of the incoming particle direction and the stopping of the particles for energy estimation respectively, resulting in different particle interaction behavior. \\

\paragraph{Particle interactions and tracking} Accelerated charged particles undergo numerous complex interactions with the matter traversed~\cite{Groom2000}. In proton imaging, charged particles are mainly influenced by Coulomb interactions with atomic electrons, decelerating the particle, as well as nuclei, randomly deflecting the particle from its straight path \cite{Gottschalk2018, Groom2000}. Additionally, on some occasions, particles undergo complex inelastic interactions with the atomic nucleus in a destructive process where the original primary particle is absorbed, and new particles are created. Due to its highly stochastic nature, secondary tracks cause additional complexities during reconstruction and are unusable for imaging. To recover usable characteristic properties of the particles, tracking algorithms aim to model or learn the pattern of the particle in the detector readouts under the influence of the inherent interaction mechanisms, aiming to reconstruct full particle trajectories.

\section{Related Work}

\paragraph{Particle tracking} While early particle tracking algorithms heavily relied on conventional algorithms such as iterative~\cite{Fruhwirth1987, Pettersen2020}, evolutionary~\cite{Mankel1997} or combinatorial \cite{Pusztaszeri1996} approaches, modern tracking solutions heavily utilize machine learning to tackle the increasing combinatorial explosion due to increasing particle counts. Especially geometric deep learning, operating either on node~\cite{Kieseler2020, Lieret2023} or edge level~\cite{DeZoort2021, Kortus2024} of graph representations, demonstrated to be highly effective. Aiming to combine advantages from conventional tracking and deep learning, recent work on RL-based tracking demonstrated both on discrete- \cite{Kortus2023} and continuous action spaces~\cite{Vage2022}, the ability to learn reconstruction policies by interacting with an environment. Our work extends the mechanisms in~\cite{Kortus2023} to a multi-agent setting.\\

\paragraph{Safe/Constrained Reinforcement Learning} Learning safe policies, operating under safety or functional constraints, is an emerging research field, both in single and multi-agent reinforcement learning.  For this work, we focus on state-wise safety by constraining the set of feasible policies. Our work is closely related to the idea of safety layers and shielding. \cite{Pham2018}, \cite{Dalal2018} and \cite{Sheebaelhamd2021}, proposed the usage of an implicit layer that performs action correction of the policy using a linearized version of the constraint function. Similarly, \cite{ElSayed2021} and \cite{Alshiekh2017} proposed the usage of safety editors, restricting  the agent to safe actions, by either reducing the safe action space or correcting unsafe actions of the policy.

\begin{figure*}[!ht]
    \centering
    \includegraphics[width=0.98\textwidth]{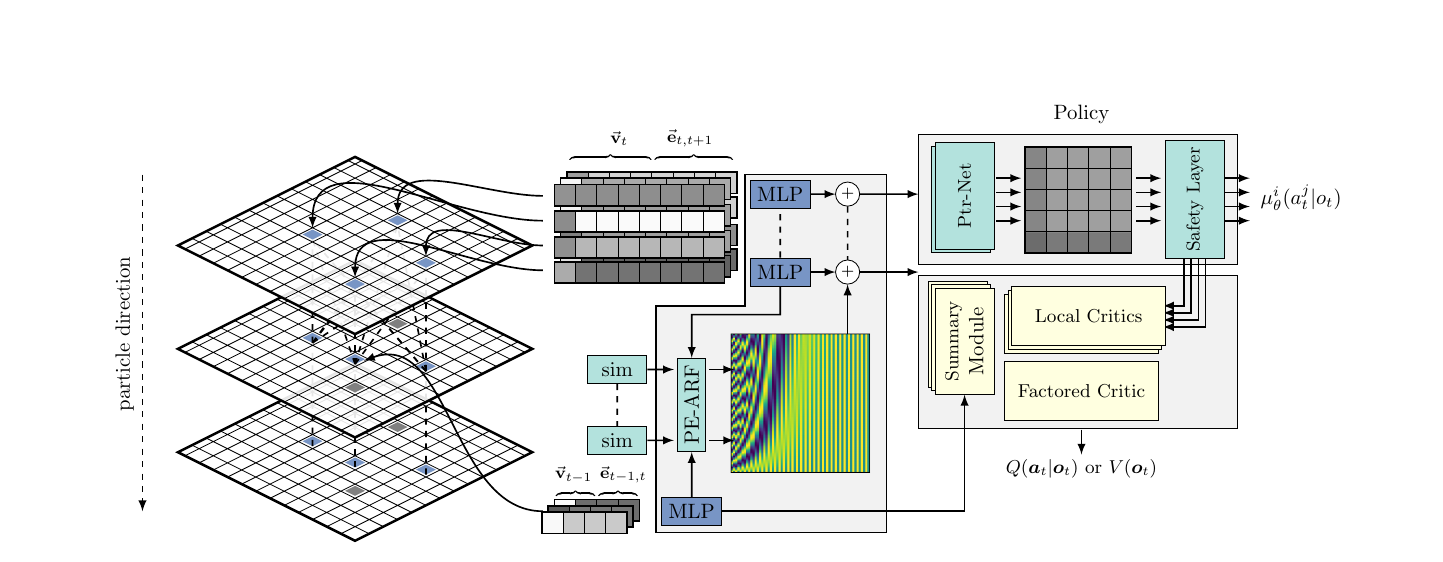}
    \caption{Interaction loop between environment description containing particle readouts in the form of a directed acyclic graph based on~\cite{Kortus2023}. The agent (network architecture on the right) observes a state, describing the current particle trajectory, and chooses a next particle hit in the subsequent layer. The reward is defined based on the physical likelihood of the undertaken transition.}
    \label{fig:rl_framework}
\end{figure*}

\section{Methodology}

In the following, we outline a general notion of constrained and unconstrained collaborative charged particle tracking, extending existing work in~\cite{Kortus2023}, and propose multiple agent architectures for the centralized training for decentralized execution (CTDE) paradigm~\cite{Oliehoek2008}.  Finally, we describe training schemes for both unconstrained and constrained MARL, highlighting the task-specific modifications and challenges.

\subsection{Problem Statement}
\label{sec:dec_pomdp}

We formulate multi-agent particle tracking over multiple layers of discrete particle readout data as a \emph{decentralized partially observable Markov decision process} (Dec-POMDP)~\cite{Bernstein2009}, operating on a directed acyclic graph. Here, $\mathcal{S}$ is a set of global (unobservable) environment states describing the current local trajectories of all agents. Instead of perceiving the global environment state, each agent can only draw individual local observations $o^{(i)}_t \sim \mathcal{O}$, defined by the last reconstructed partial track segment, and all possible next segments $ o_t^{(i)} = \{v_{t}, e_{t-1, t}\} \cup \left\{v_{t+1}^{(i)}, e_{t, t+1}^{(i)}\right\}.$ Each agent can select, based on its perceived observation, from a set of actions defined by the set of next hit candidates, which we treat later on either as unconstrained or constrained (by unique assignments). For each interaction, all agents receive a scalar reward signal $r_t$, accumulated until a terminal state triggered by the absence of a valid action as the end of the detector, is reached. \\

\paragraph{Graph construction} Following the parametrization of particle readouts described in \cite{Kortus2023}, we model the particle data, as a directed acyclic graph~(\emph{hit graph}), where each hit represents a vertex in the graph. Edges are generated between hits of adjacent layers, opposite to the direction of the particle. Both, vertices and edges are parametrized by a set of features $\boldsymbol{v}_i = [\Delta E, x, y, \mathbbm{1}_z]$ and $\boldsymbol{e}_{ij} = [r_{ij}, \theta_{ij}, \phi_{ij}]$, defining the energy deposition and position of the hit with one-hot encoded layer index as well as the spherical coordinates of the edge connections. Finally, we employ the feature normalization scheme of~\cite{Kortus2023}, compensating for the beam position in the detector, providing translation invariant features. \\

\paragraph{Sampling of track candidates} Track candidates are constructed for a hit graph, starting from all initial unoccupied graph vertices in the last detector layer, by iteratively adding new vertices in subsequent layers, until a terminal state is reached. Unassigned vertices in subsequent layers are incrementally added to the list of track candidates. To provide a starting track segment, functioning as an initial local observation, we rely on ground-truth seeding~\cite{Kortus2023}, avoiding unwanted dependencies of seeding algorithms on the performance of the proposed algorithms and providing a performance upper bound of RL-based tracking.  \\

\paragraph{Objective} We attempt to find, by repeatedly interacting in the described environment, generating sampled track candidates, a joint policy, that collaboratively maximizes the gathered expected discounted return under a shared team reward.  Similar to \cite{Kortus2023}, we aim to optimize the reconstruction policy by minimizing the average amount of particle scattering in a readout frame over all agents. We thus define the reward signal as the negative average scatter angle obtained for each transition in the graph. In the multi-agent case, we rely on this naive description over the more detailed modelling of the energy dependent scattering behavior~\cite{Highland1975}, described in~\cite{Kortus2023}, to remove the dependence of the reward signal on full track candidates, making it more suitable for off-policy algorithms.

\subsection{Architecture and Implementation}
\label{sec:architecture}

In this section, we describe extensions to the existing attention-based agent parametrization~\cite{Kortus2023}, providing both a permutation invariant and action size independent processing. Our main focus lies on centralized critic components, that can be seamlessly integrated into the existing framework for particle tracking~\cite{Kortus2023}. To improve over the existing architecture, we simplify the policy by moving computationally intensive layers from the policy to the centralized critic. Finally, we propose the use of a differentiable safety layer, similar to \cite{Dalal2018, Sheebaelhamd2021} for constrained particle tracking, guaranteeing unique assignments of particle hits. We further provide useful gradient information, building upon existing work in decision-focused learning by~\cite{Vlastelica2020, Sahoo2023}. \\

\paragraph{Feature preparation} Following the description of local observations in Section~\ref{sec:dec_pomdp}, we extract edge- and node-level features, for both last reconstructed ($v_{t-1}\rightarrow v_t$) and possible next track segments ($v_{t}\rightarrow v_{t+1,j}$) from the hit graph according to

\begin{equation}
\begin{split}
    \boldsymbol{h}^{(i)}_{obs} &= \Psi_1\left([\boldsymbol{v}_t, \boldsymbol{e}_{t-1,t}]\right) \quad \text{and} \\  \boldsymbol{h}^{(i)}_{act,j} &= \Psi_2\left([\boldsymbol{v}_{t+1}, \boldsymbol{e}_{t,t+1,j}]\right),
\end{split}
\end{equation}

which are projected by separate multi-layer perceptrons into an equally sized higher dimensional feature space. For performance reasons, we omit the additional feature vector generated by a graph neural network as proposed in~\cite{Kortus2023}, as we found the simple feature description to be sufficient in combination with the use of a safety layer. The \emph{positional encoding with adaptive receptive field} (PE-ARF) mechanism, proposed in~\cite{Kortus2023}, is used to provide additional positional information in the form of cosine similarities restricted to a learnable area of interest.  \\

\paragraph{Local agent policies} We parameterize the local policy~$\mu^{(i)}_\theta$ of each agent using a pointer mechanism~\cite{Vinyals2015} (Ptr-Net), predicting the conditional probability of the local action $a^{(i)}_{t,j}$ conditioned on \emph{observation-} and \emph{action features}. This mechanism is defined by additive attention~\cite{Bahdanau2015} according to
\begin{equation}
    \alpha^{(i)}_{j,t} = \boldsymbol{v}^T \tanh(\boldsymbol{W}_1\boldsymbol{h}^{emb}_{act,ij}  + \boldsymbol{W}_2\boldsymbol{h}^{emb}_{obs,i}),
\end{equation}

where $\boldsymbol{W}_1$, $\boldsymbol{W}_2$ and $\boldsymbol{v}$ are learnable parameter matrices/vectors. The output scorings are normalized over all possible segments using a softmax activation.  \\

\paragraph{Communication}  We focus in this work on decentralized actor architectures, requiring no or minimal global communication during inference, thus minimizing the computational overhead of communication protocols. While~\cite{Kortus2023} uses multi-head attention (MHA) to learn an agreement between segment candidates, we consider this mechanism as a form of centralization and thus reallocate it from the actor to the centralized critic for all multi-agent architectures, reducing the computational cost of evaluating the policy.\\

\paragraph{Safety Policy Layer} To correct the predicted local policies for duplicate assignments, we propose, similar to~\cite{Kortus2024}, the usage of a centralized safety layer \cite{Dalal2018, Sheebaelhamd2021}, performing for every reconstruction step an action correction for the learned joint policy by solving a \emph{linear sum assignment problem} (LSAP). The safety layer ensures during both training and inference a full or partial unique matching defined by 
\begin{equation}
    \begin{split}
        \min &\quad \sum_{(i,j) \in \mathcal{E}} \widehat{\mu}_{ij}c_{ij} \\
        \text{s.t.} &\quad \sum_{i \in \mathcal{V}_S} \widehat{\mu}_{ij} = 1, \quad j \in \mathcal{V}_T \\
        &\quad \sum_{j\in \mathcal{V}_T} \widehat{\mu}_{ij} \leq 1, \quad i \in \mathcal{V}_S \\
    \end{split},
\end{equation}

that minimizes the required cost of deviating from the proposed local policies. Here $c_{ij} \in \hat{\boldsymbol{\mathcal{C}}}$ are the individual elements of a $n\times m$ cost  matrix, defined, either by infinite cost for assignments already occupied by another track due to its initial seeding mechanism, or by the L2-norm of the local policy to the one-hot encoding of the corresponding target vertex, according to

\begin{equation}
    c_{ij} = \begin{cases} 
    \Vert \boldsymbol{\mu}^{i}(a_{j} \vert \boldsymbol{o}) - \mathbbm{1}(a_{j})\Vert_2^2 & \text{if not used for seeding}\\
    \infty & \text{otherwise}.
    \end{cases}
\end{equation}

By projecting the unsafe action, the action-corrected policy becomes inherently deterministic, requiring off-policy optimization and an exploratory policy for generating training samples. We sample track candidates with random exploration using parameter noise~\cite{Fortunato2018, Plappert2018}. We therefore replace the linear layers of the pointer mechanism with noisy linear layers \cite{Fortunato2018}. While~\cite{Dalal2018} and \cite{Sheebaelhamd2021} propose a safety layer, that performs action correction without being able to differentiate through the layer itself, we use blackbox gradient information to reduce the complexity of the learning task, especially for the high dimensionality of the assignment problem. \\

\paragraph{Blackbox differentiation} To provide gradient information for a combinatorial solver of the general form $y(\boldsymbol{\mathcal{C}}) = \arg\min_{y \in \mathcal{Y}} c(\boldsymbol{\mathcal{C}}, y)$, \cite{Vlastelica2020} proposed, substituting the piecewise constant solvers mapping of combinatorial solvers at the point $\hat{\boldsymbol{\mathcal{C}}}$ by a linear interpolation between the points $\hat{\boldsymbol{\mathcal{C}}}$ and $\boldsymbol{\mathcal{C}}'$ according to
\begin{equation}
    \nabla^{BB}_{\boldsymbol{\mathcal{C}}} f_\lambda (\hat{\boldsymbol{\mathcal{C}}}) := - \frac{1}{\lambda} \left[ y(\hat{\boldsymbol{\mathcal{C}}}) - y_\lambda(\boldsymbol{\mathcal{C}}') \right], \quad \text{where}
\end{equation}

\begin{equation}
    \quad \boldsymbol{\mathcal{C}}' = \text{clip}\left(\hat{\boldsymbol{\mathcal{C}}} + \lambda \frac{dL}{dy}\left(y(\hat{\boldsymbol{\mathcal{C}}})\right), 0, \infty \right).
\end{equation}

Here, $y(\hat{\boldsymbol{\mathcal{C}}})$ and $y(\boldsymbol{\mathcal{C}}')$ are solutions generated by predicted and perturbed cost.  Further, $\lambda \in \mathbb{R}^+$ functions as a tunable hyperparameter, interpolating between truthfulness and informativeness of the gradients~\cite{Vlastelica2020}. The usefulness of the gradient information for particle tracking has been already demonstrated in~\cite{Kortus2024}. \\

\paragraph{Cost margins}  With increasing number of solution sets, the policy becomes prone to settle changes in the cost matrix, limiting generalization. \cite{Sahoo2023} proposed, adding random noise to the predicted cost, increasing the margin to the decision boundaries of the predictive output. As we found this mechanism to be highly instable for our use case, we instead add an additional component $\nabla^{\leftrightarrow}_w$ to the BB-scheme, with

 \begin{equation}
    \nabla^{BB}_{\boldsymbol{\mathcal{C}}} f_\lambda (\hat{\boldsymbol{\mathcal{C}}}) + \nu \nabla^{\leftrightarrow}_{\boldsymbol{\mathcal{C}}} f(\hat{\boldsymbol{\mathcal{C}}}), \quad \text{where}\quad \nabla^{\leftrightarrow}_{\boldsymbol{\mathcal{C}}} f(\hat{\boldsymbol{\mathcal{C}}}) = y(\hat{\boldsymbol{\mathcal{C}}}),
 \end{equation}

forcing the assignments of the joint policy $\boldsymbol{\mu}$ in the direction of lower assignment costs. The influence of $\nabla^{\leftrightarrow}_{\boldsymbol{\mathcal{C}}}$ can be controlled using the hyperparameter $\nu \in \mathbb{R}^+$. \\
    
\paragraph{Centralized critic}  To mitigate instationarity, introduced by the otherwise independent learners~\cite{Tan1993, Sunehag2018}, we propose centralized factored critic functions  for state- $V^\theta(\boldsymbol{o}_t)$ and action-value function $Q^\theta(\boldsymbol{a}_t \vert \boldsymbol{o}_t)$, decomposing the global value function into agent-wise values~\cite{Sunehag2018} according to
\begin{equation}
Q(\boldsymbol{a}_t, \boldsymbol{o}_t) \approx \frac{1}{N} \sum_{i=1}^{N} Q^{(i)}_\theta\left(a_{t}^{(i)}, o_{t}^{(i)}, \phi(\boldsymbol{o}_t)\right) 
\end{equation}

\begin{equation}
    \quad V(\boldsymbol{o}_t) \approx \frac{1}{N} \sum_{i=1}^{N} V^{(i)}_\theta\left(o_{t}^{(i)}, \phi(\boldsymbol{o}_t)\right).
\end{equation}

Each agent-wise value is composed, using local and global information, utilizing a mixture of additive~\cite{Bahdanau2015} and self-attention~\cite{Iqbal2019}. To provide for each agent a single feature, we compress the set of agent observations $\langle \boldsymbol{h}_{obs}, \boldsymbol{h}^{(1)}_{act}, \dots, \boldsymbol{h}^{(N)}_{act} \rangle$ for both $V^\theta$ and $Q^\theta$. For the action dependent Q-function, we model the compressed representation $\boldsymbol{h}_{Q}^{(i)}$ by a joint policy weighted function of observation- action features according to

\begin{equation}
        \boldsymbol{h}_{Q}^{(i)} = \left[\sum_{j=1}^{M} \boldsymbol{\mu}^i(\boldsymbol{a}_{t,j}, \boldsymbol{o}_t)\left(\boldsymbol{h}_{\text{obs,i}}^{emb,(i)} +  \overline{\boldsymbol{h}}_{\text{act},j}^{emb,(i)}\right) \right].
\end{equation}

Here, $\overline{h}$ is an assembled feature over true and uncorrelated reference action features aggregated as a weighted sum over multiple random samples from a replay buffer $\mathcal{D}$ following

\begin{equation}
    \overline{\boldsymbol{h}}_{act,j}^{emb,(i)} = \boldsymbol{h}_{act,j}^{emb,(i)} + \gamma \sum_{t',i',j'\sim \mathcal{D}} \boldsymbol{h}_{act,j',t'}^{emb,(i')},
\end{equation}

where $\gamma$ is a hyperparameter. This expression functions as a smoothing and regularization term with contextual information, allowing for reduced variance during training, improving convergence.  For the action-independent state-value function $V^\mu_\theta$, the weighting of the action features is replaced by a learnable weighting, modelled using an additive attention mechanism~\cite{Bahdanau2015} according to
\begin{equation}
        \boldsymbol{h}_{V}^{(i)} = \left[\boldsymbol{h}_{\text{obs}}^{(i)} + \sum_{j=1}^{M} \alpha_{j} \boldsymbol{h}_{\text{act},j}^{(i)} \right], \quad \text{with}
\end{equation}

\begin{equation}
    \alpha^{(i)}_j = \boldsymbol{v}^T \tanh\left(\boldsymbol{W}_1\boldsymbol{h}^{emb,(i)}_{act,j}  + \boldsymbol{W}_2\boldsymbol{h}^{emb,(i)}_{obs}\right).
\end{equation}

The soft weighting makes the cross-state regularization for variance reduction obsolete. Further, we encourage global communication between agents in form of two stacked self-attention blocks with layer normalization~\cite{Ba2016} and skip connections~\cite{He2016}, each defined as
\begin{equation}
    \label{eq:mha1}
    \boldsymbol{h}_{Q/V}^{(i,l)} = \text{LN}\left(\boldsymbol{h}_{Q/V}^{(i, l-1)} + \text{ReLU}\left(\text{MHA}\left(\boldsymbol{h}_{Q/V}^{(1:N, l-1)}
    \right)\right)\right)
\end{equation}

Finally, factored values are obtained as the average agent-wise estimate conditioned on $\boldsymbol{h}_{Q}^{(i)}$/$\boldsymbol{h}_{V}^{(i)}$ using an MLP. The value range for $Q$ and $V$ is restricted for either raw- (sigmoid) or normalized rewards (tanh) accordingly (additional details in Section~\ref{sec:optim}) and scaled by the learnable parameter $s$.
\begin{equation}
       Q(s, a) = - \frac{1}{N} \sum_{i=1}^{N}  s \cdot \sigma\left(\Phi_{Q}\left(\boldsymbol{h}^{(i)}_{Q}\right)\right)
\end{equation}

 \begin{equation}
       V(s) = - \frac{1}{N} \sum_{i=1}^{N}  s \cdot \tanh\left(\Phi_{V}\left(\boldsymbol{h}^{(i)}_{V}\right)\right).
\end{equation}

For completed particle tracks without valid assignments~(early termination), we employ a value masking, where the relevant local agent-wise value estimates are excluded from the global value estimate. This representation prevents the observation of rewards obtained after early termination, posing additional complexity to the credit assignments~\cite{Cohen2021}, however, we choose the masking mechanism in favor of simplicity of the overall architecture\footnote{While we didn't witness significant issues in credit assignment, incremental updates of the architecture could introduce absorbing states for agents with early termination~\cite{Cohen2021}, potentially further improving the learning abilities.}.

\begin{figure*}[ht!]
    \centering
    \includegraphics[width=\textwidth]{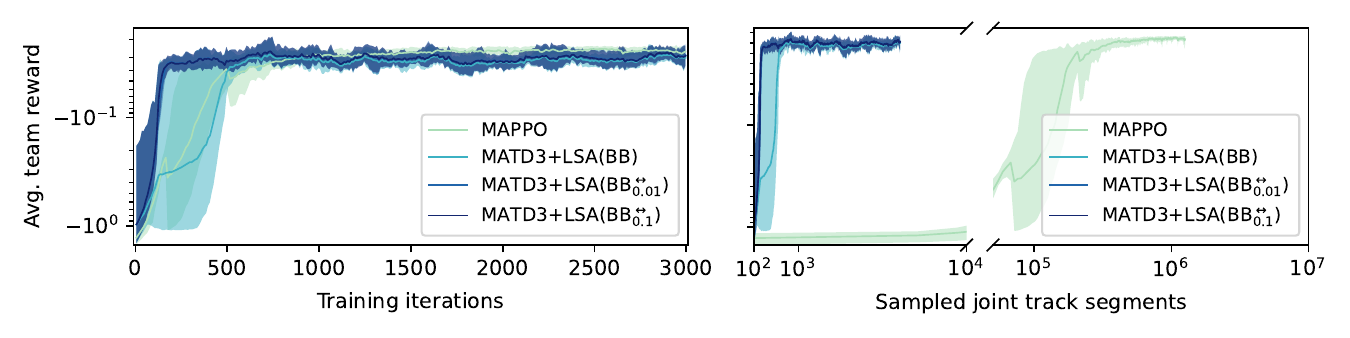}
    \caption{Average return obtained by the agents over time during training, plotted as a function of performed updates (for MAPPO: iteration over all epochs are counted as a single update) and sampled track transitions.}
    \label{fig:rewards}
\end{figure*}

\subsection{Optimization of Agents}
\label{sec:optim}

The following section outlines the different optimization schemes for optimizing both unconstrained and constrained agents. Here we put specific focus on the details and modifications required for particle tracking.\\

\paragraph{Unconstrained on-policy baseline} We optimize an unconstrained joint policy using multi-agent proximal policy optimization algorithm (MAPPO)~\cite{Yu2022, Ma2022}, providing an extrapolation of the learning abilities of~\cite{Kortus2023} to a collaborative multi-agent setting. We use the architecture described in~Section~\ref{sec:architecture}, replacing the deterministic joint policy $\mu_\theta$ by an unconstrained stochastic policy $\pi_\theta$ and a centralized state-value estimator $V_{\boldsymbol{\pi}}^\theta$. We estimate team advantages using the generalized advantage estimator~\cite{Schulman2016} and employ independent reward normalization for calorimeter and tracker layer, following the normalization scheme in~\cite{Kortus2023}. \\

\paragraph{Off-policy optimization} To cope with the deterministic safety-layer corrected policies, we optimize it similarly to~\cite{Sheebaelhamd2021}, using a multi-agent variant of the \emph{Deep Deterministic Policy Gradient} (DDPG) algorithm~\cite{Lillicrap2016}. However, while~\cite{Sheebaelhamd2021} uses the multi-agent DDPG algorithm \cite{Lowe2017}, we found the MATD3~\cite{Ackermann2019} algorithm with two critic networks, mitigating overestimation bias, together with periodical hard critic updates worked superior for our use case. We found for the independent reward normalization mechanism to have a negative impact on optimization. Finally, we use a replay buffer with a small buffer size, owed to the quickly changing distribution of samples of the large joint action space~\cite{Hu2021}.

\section{Experiments}
\label{sec:experiments}

\begin{table}
\caption{Overview of all considered RL and MARL particle tracking schemes evaluated in Section~\ref{sec:experiments} }
\label{tab:experiments}
\centering
\resizebox{\linewidth}{!}{%
\begin{NiceTabular}{llccccl}
\toprule
Name & Alg. &   Centr. V/Q & $\gamma$ & SL(T)  & SL(E)& SL-grad. \\
\midrule 
PPO                                             & \cite{Schulman2017}   &            &  &            &             &                         \\ 
PPO+LSA                                         & \cite{Schulman2017}   &            &  &            & \checkmark  &                         \\ 
MAPPO                                           & \cite{Yu2022}         &\checkmark  &  &  &             &                         \\
MATD3+LSA (BB)                                  & \cite{Ackermann2019}  &\checkmark  & 0.75 & \checkmark & \checkmark  & BB\cite{Vlastelica2020} \\
MATD3+LSA (BB$^\leftrightarrow_\nu$)            & \cite{Ackermann2019}  &\checkmark  & 0.25 & \checkmark & \checkmark  & BB + ours \\
\bottomrule
\end{NiceTabular}}
\end{table}

For the studies reported in this work, we rely on Monte-Carlo~(MC) simulations of detector readout data~\cite{Kortus2022}, generated using the GATE toolkit~\cite{Jan2004, Jan2011} based on the Geant4 simulation framework~\cite{Agostinelli2003, Allison2006, Allison2016}. The dataset consists of multiple simulations with and without water phantom (\SI{100}{mm}, \SI{150}{mm} and \SI{200}{mm}), positioned between the particle beam and detector. The data is further diversified by manually splitting the data into readout frames of different particle densities ($p^+/F$) of 50, 100, 150 and 200. Each simulation consists of  10,000 simulated primary particles. All data is publicly available on Zenodo~\cite{Kortus2022}. \\

\paragraph{Configurations} To explore the performance of single- and multi-agent systems of various degrees of complexity, we construct variations of the agent described in the previous sections, summarized in Table~\ref{tab:experiments}. Each variant is constructed based on the selected optimization algorithm, the usage of a safety layer (during training SL(T) and execution SL(E)) as well as the differentiation scheme. We couldn't find a stable MATD3 configuration without a safety layer that consistently converged to low-reward solutions, and thus excluded it from the results. The single agent results for PPO and PPO+LSA are based on the trained models in~\cite{Kortus2023}.  \\

\paragraph{Training procedure}  We use particle simulations without any absorber material between beam source and detector for optimization, providing a worst-case scenario in terms of secondary production and track length. We then train, for each configuration in Table~\ref{tab:experiments}, five independent policies on sampled track candidates with a particle density of 50 primary particles per readout frame, to obtain robust results with confidence intervals.   \\

\paragraph{Baselines} In addition to the multi-agent schemes, listed in Table~\ref{tab:experiments}, we compare the reconstruction performance, with both two single-agent variants of particle tracking described in~\cite{Kortus2023} (with an additional centralized version using the proposed safety layer during inference) and a sequential track follower searching for solutions that minimize the total amount of scattering~\cite{Pettersen2020}. To obtain comparable results, all techniques construct the initial seed used for tracking using ground-truth information.  \\

\paragraph{Performance metrics} We assess and compare the performance of the proposed tracking algorithms using track purity~($p$) and efficiency~($\epsilon$), estimated after prior rejecting partial or implausible tracks using simple cuts for scattering angle and energy deposition according to~\cite{Pettersen2021}. For assessing the correctness of a track, we rely on a \emph{perfect matching} criterion, where all hits in a track need to be correctly assigned.

\begin{figure}
    \centering
    \includegraphics[width=\linewidth]{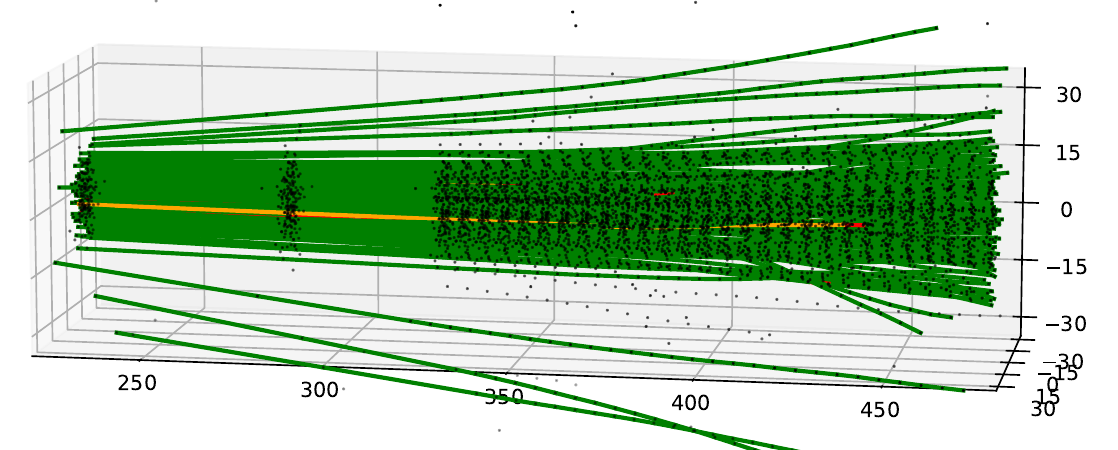}
    \caption{Particle tracks generated using MATD3+LSA (BB$^\leftrightarrow_{\nu=0.01}$) for simulated particle tracks with 100mm water phantom and 200$p^+/F$ }
    \label{fig:enter-label}
\end{figure}

\begin{table*}[!ht]
\caption{Reconstruction performance for water phantoms of 100, 150 and 200~mm thickness and 100, 150 and 200 $p^{+}/F$. Results for PPO and Track follower are taken from \cite{Kortus2023}. Elements marked with hatched lines are outside the range of the colormap.}
\label{tab:results_all}
\centering
\resizebox{\textwidth}{!}{
\scriptsize
\begin{NiceTabular}{l|l|cc|cc|cc}
\toprule
    &       & \multicolumn{2}{l}{\textbf{100~mm Water Phantom}} & \multicolumn{2}{l}{\textbf{150~mm Water Phantom}} & \multicolumn{2}{l}{\textbf{200~mm Water Phantom}} \\
\midrule
$p^+/F$ & Algorithm & $p$ [\%]      & $\epsilon$ [\%] & $p$ [\%] & $\epsilon$ [\%]& $p$ [\%]    &$\epsilon$ [\%] \\
\midrule
50 & Track follower \cite{Pettersen2020}            & \gradient{88.1}{0.0}  & \gradient{79.7}{0.0}  & \gradient{90.3}{0.0}  & \gradient{82.7}{0.0}  & \gradient{91.2}{0.0}  & \gradient{83.8}{0.0}\\
    & PPO \cite{Kortus2023}                         & \gradient{92.5}{0.2}  & \gradient{81.5}{0.3}  & \gradient{93.8}{0.1}  & \gradient{84.0}{0.4}  & \gradient{94.5}{0.1}  & \gradient{85.5}{0.2}\\
    & MAPPO                                         & \gradient{80.1}{21.7} & \gradient{70.3}{19.5} & \gradient{82.7}{19.5} & \gradient{73.6}{18.0} & \gradient{83.9}{19.7} & \gradient{75.8}{18.0} \\
    & MATD3+LSA (BB)                                & \gradient{56.6}{21.5} & \gradient{48.6}{19.3} & \gradient{63.5}{22.4} & \gradient{55.2}{21.1} & \gradient{68.8}{22.9} & \gradient{60.1}{22.9} \\
    & MATD3+LSA (BB$^\leftrightarrow_{\nu=0.01}$)   & \gradient{96.2}{0.1}  & \textbf{\gradient{84.0}{0.1}}  & \textbf{\gradient{97.0}{0.1} } & \textbf{\gradient{85.9}{0.1}}  & \textbf{\gradient{97.3}{0.1}}  & \textbf{\gradient{87.3}{0.0}}\\
    & MATD3+LSA (BB$^\leftrightarrow_{\nu=0.1}$\;)  & \textbf{\gradient{96.3}{0.2}}  & \textbf{\gradient{84.0}{0.2}}  & \gradient{96.9}{0.1}  & \gradient{85.7}{0.1}  & \textbf{\gradient{97.3}{0.1}}  & \gradient{87.2}{0.2}\\
\midrule
100 & Track follower \cite{Pettersen2020}           & \gradient{83.0}{0.0}  & \gradient{74.6}{0.0}  & \gradient{86.6}{0.0}  & \gradient{79.0}{0.0}  & \gradient{87.4}{0.0}  & \gradient{80.3}{0.0}\\
    & PPO \cite{Kortus2023}                         & \gradient{85.7}{0.2}  & \gradient{75.1}{0.5}  & \gradient{89.0}{0.2}  & \gradient{79.1}{0.5}  & \gradient{89.5}{0.1}  & \gradient{80.9}{0.3}\\
    & MAPPO                                         & \gradient{71.3}{24.1} & \gradient{62.4}{21.5} & \gradient{75.1}{23.0} & \gradient{66.3}{21.3} & \gradient{76.3}{23.3} & \gradient{68.8}{21.2} \\
    & MATD3+LSA (BB)                                & \gradient{40.2}{20.4} & \gradient{34.2}{17.6} & \gradient{48.6}{23.2} & \gradient{42.0}{20.8} & \gradient{55.0}{25.3} & \gradient{48.1}{23.4} \\
    & MATD3+LSA (BB$^\leftrightarrow_{\nu=0.01}$)   & \textbf{\gradient{91.9}{0.2}}  & \textbf{\gradient{79.5}{0.2}}  & \textbf{\gradient{94.1}{0.1}}  & \textbf{\gradient{82.5}{0.2}}  & \gradient{93.6}{0.1}  & \textbf{\gradient{83.5}{0.1}}\\
    & MATD3+LSA (BB$^\leftrightarrow_{\nu=0.1}$\;)  & \textbf{\gradient{91.9}{0.2}}  & \textbf{\gradient{79.5}{0.2}}  & \gradient{94.0}{0.2}  & \gradient{82.4}{0.2}  & \textbf{\gradient{93.7}{0.1}}  & \textbf{\gradient{83.5}{0.2}}\\
\midrule
150 & Track follower \cite{Pettersen2020}           & \gradient{79.1}{0.0}  & \gradient{70.9}{0.0}  & \gradient{83.2}{0.0}  & \gradient{75.7}{0.0}  & \gradient{84.7}{0.0}  & \gradient{77.7}{0.0}\\
    & PPO \cite{Kortus2023}                         & \gradient{80.6}{0.3}  & \gradient{70.8}{0.6}  & \gradient{84.0}{0.1}  & \gradient{74.5}{0.6}  & \gradient{85.5}{0.2}  & \gradient{77.1}{0.3}\\
    & MAPPO                                         & \gradient{65.0}{24.4} & \gradient{57.2}{21.8} & \gradient{69.4}{23.4} & \gradient{61.3}{21.9} & \gradient{71.3}{24.4} & \gradient{64.3}{22.2} \\
    & MATD3+LSA (BB)                                & \gradient{31.4}{18.0} & \gradient{26.6}{15.3} & \gradient{39.6}{21.5} & \gradient{34.0}{18.9} & \gradient{46.4}{24.4} & \gradient{40.6}{22.0} \\
    & MATD3+LSA(BB$^\leftrightarrow_{\nu=0.01}$)    & \textbf{\gradient{88.8}{0.2}}  & \textbf{\gradient{76.8}{0.3}}  & \gradient{90.9}{0.2}  & \textbf{\gradient{79.2}{0.2}}  & \gradient{91.2}{0.2}  & \gradient{81.1}{0.2}\\
    & MATD3+LSA(BB$^\leftrightarrow_{\nu=0.1}$\;)   & \textbf{\gradient{88.8}{0.4}}  & \gradient{76.7}{0.4}  & \textbf{\gradient{91.1}{0.3}}  & \textbf{\gradient{79.2}{0.3}}  & \textbf{\gradient{91.4}{0.2}}  & \textbf{\gradient{81.2}{0.3}}\\
\midrule
200 & Track follower \cite{Pettersen2020}           & \gradient{75.4}{0.0}  & \gradient{67.4}{0.0}  & \gradient{80.1}{0.0}  & \gradient{72.9}{0.0}  & \gradient{81.6}{0.0}  & \gradient{75.0}{0.0}\\
    & PPO \cite{Kortus2023}                         & \gradient{75.5}{0.3}  & \gradient{66.6}{0.6}  & \gradient{80.3}{0.4}  & \gradient{71.1}{0.6}  & \gradient{81.9}{0.3}  & \gradient{73.9}{0.4}\\
    & MAPPO                                         & \gradient{59.6}{23.6} & \gradient{52.8}{21.2} & \gradient{65.2}{23.5} & \gradient{57.6}{22.0} & \gradient{66.9}{24.8} & \gradient{60.5}{22.6} \\
    & MATD3+LSA(BB)                                 & \gradient{25.8}{15.8} & \gradient{21.8}{13.3} & \gradient{33.7}{19.4} & \gradient{28.9}{16.8} & \gradient{40.7}{22.7} & \gradient{35.6}{20.3} \\
    & MATD3+LSA (BB$^\leftrightarrow_{\nu=0.01}$)   & \gradient{84.7}{0.3}  & \gradient{73.0}{0.3}  & \gradient{88.2}{0.2}  & \gradient{76.6}{0.3}  & \gradient{88.2}{0.2}  & \gradient{78.2}{0.2}\\ 
    & MATD3+LSA (BB$^\leftrightarrow_{\nu=0.1}$\;)  & \textbf{\gradient{84.9}{0.3}}  & \textbf{\gradient{73.3}{0.3}}  & \textbf{\gradient{88.6}{0.3}}  & \textbf{\gradient{76.7}{0.3}}  & \textbf{\gradient{88.4}{0.3}}  & \textbf{\gradient{78.3}{0.4}}\\
\bottomrule
\end{NiceTabular}}
\end{table*}

\begin{table*}[!ht]
\caption{Reconstruction performance, measured in terms of purity $p$ and efficiency $\epsilon$ for water phantoms of 100, 150 and 200~mm thickness and 100, 150 and 200 $p^{+}/F$. Results for PPO+LSA are generated with the models from~\cite{Kortus2023}}
\label{tab:results_lsa}
\centering
\resizebox{\textwidth}{!}{
\scriptsize
\begin{NiceTabular}{l|l|cc|cc|cc}
\toprule
    &       & \multicolumn{2}{l}{\textbf{100~mm Water Phantom}} & \multicolumn{2}{l}{\textbf{150~mm Water Phantom}} & \multicolumn{2}{l}{\textbf{200~mm Water Phantom}} \\
\midrule
$p^+/F$ & Algorithm & $p$ [\%]      & $\epsilon$ [\%] & $p$ [\%] & $\epsilon$ [\%]& $p$ [\%]    &$\epsilon$ [\%] \\
\midrule
50  & MATD3+LSA (BB$^\leftrightarrow_{\nu=0.1}$)            & \textbf{\gradient{96.3}{0.2}} & \textbf{\gradient{84.0}{0.2}} & \gradient{96.9}{0.1} & \textbf{\gradient{85.7}{0.1}} & \textbf{\gradient{97.3}{0.1}} & \textbf{\gradient{87.2}{0.2}}\\
    & PPO+LSA                                               & \gradient{95.9}{0.2} & \gradient{83.3}{0.6} & \textbf{\gradient{97.0}{0.1}} & \textbf{\gradient{85.7}{0.4}} & \gradient{97.2}{0.3} & \textbf{\gradient{87.2}{0.4}} \\
\midrule
100 & MATD3+LSA (BB$^\leftrightarrow_{\nu=0.1}$)            & \textbf{\gradient{91.9}{0.2}} & \textbf{\gradient{79.5}{0.2}} & \textbf{\gradient{94.0}{0.2}} & \textbf{\gradient{82.4}{0.2}} & \textbf{\gradient{93.7}{0.1}} & \textbf{\gradient{83.5}{0.2}}\\
    & PPO+LSA                                               & \gradient{91.5}{0.4} & \gradient{79.0}{0.5} & \textbf{\gradient{94.0}{0.2}} & \gradient{82.3}{0.3} & \gradient{93.6}{0.4} & \gradient{83.3}{0.4} \\
\midrule
150 & MATD3+LSA(BB$^\leftrightarrow_{\nu=0.1}$)             & \textbf{\gradient{88.8}{0.4}} & \textbf{\gradient{76.7}{0.4}} & \textbf{\gradient{91.1}{0.3}} & \textbf{\gradient{79.2}{0.3}} & \textbf{\gradient{91.4}{0.2}} & \textbf{\gradient{81.2}{0.3}}\\
    & PPO+LSA                                               & \gradient{88.4}{0.4} & \gradient{75.9}{0.9} & \gradient{90.5}{0.4} & \gradient{78.6}{0.6} & \gradient{90.8}{0.5} & \gradient{80.2}{0.5} \\
\midrule
200 & MATD3+LSA (BB$^\leftrightarrow_{\nu=0.1}$)            & \textbf{\gradient{84.9}{0.3}} & \textbf{\gradient{73.3}{0.3}} & \textbf{\gradient{88.6}{0.3}} & \textbf{\gradient{76.7}{0.3}} & \textbf{\gradient{88.4}{0.3}} & \textbf{\gradient{78.3}{0.4}}\\
    & PPO+LSA                                               & \gradient{84.0}{0.5} & \gradient{72.0}{0.9} & \gradient{87.9}{0.4} & \gradient{75.8}{0.7} & \gradient{87.7}{0.8} & \gradient{77.1}{0.6} \\
\bottomrule
\end{NiceTabular}
}
\end{table*}

\subsection{Optimization and Tracking Performance}
\label{sec:results}

We examine and compare the performance for all configurations in Table~\ref{tab:experiments} to identity and quantify the necessary factors for multi-agent based particle tracking using MARL. Figure~\ref{fig:rewards} shows the average reward obtained as a function of network updates and sampled track segments. Here, we find similar training performance for the on-policy MAPPO and off-policy MATD3 approaches for equal number of training iterations. However, due to the on-policy nature of MAPPO, requiring data generated from the current policy, this approach requires significantly more transitions to converge and is thus significantly more sample inefficient than the off-policy MATD3 algorithm, utilizing a replay buffer. Further, while all multi-agent variants except for the unconstrained MATD3 approach, which we excluded from the experiments, converge to high average team rewards, MAPPO converges consistently to the highest average reward, suggesting the best optimization behavior of all. Finally, we find that both constrained agents with cost margins show significantly faster convergence to high rewards, requiring approximately 300 training iterations less than the other agents.

Table~\ref{tab:results_all} summarizes the reconstruction performance (purity $p$ and efficiency $\epsilon$) of all MARL and baseline algorithms. We find that, while achieving lower average rewards compared to MAPPO, MATD3+LSA (BB$^\leftrightarrow_{\nu}$\;) outperforms all baseline and MARL variants in both configurations of $\nu$ by a significant margin. Especially for higher particle densities, the constrained policy with cost margins can benefit from the increased assignment complexity, outperforming the single-agent and unconstrained algorithms. We find the safety layer to be a critical component in multi-agent tracking, allowing for efficient sampling during training and inference, simplifying spacial credit assignment across agents, while avoiding duplicate assignment of particle hits. Further, we find the performance of MATD3+LSA(BB$^\leftrightarrow_\nu$) to be robust to exact choice of $\nu$, producing similar results for both selected configurations.

To quantify the impact of the multi-agent optimization, we compare the performance of MATD3+LSA(BB$^\leftrightarrow_{\nu=0.1}$\;) with a post-training centralized version of the single-agent PPO algorithm (PPO+LSA). Table~\ref{tab:results_lsa} shows that PPO+LSA achieves similar performance, with only slight improvements in performance for the multi-agent approach. We find that the overall difference in performance is statistically not or only marginally significant (avg. p-values obtained by one-sided ttest \cite{Welch1947}: $p$: 0.19, $\epsilon$: 0.12), demonstrating the strong ability of single-agent RL to efficiently learn reasonable conditional probabilities usable to resolve assignment conflicts during inference. Similar results are presented in~\cite{Kortus2024} for supervised learning. However, for large particle multiplicities (e.g. 200 $p^+/F$) we find the constrained multi-agent approach to outperform the single-agent approach by $0.75$ percentage points (pp) (p-value: 0.03) in purity and $1.12$ pp (p-value: 0.02) in efficiency, while only using limited information of the single-agent reward, indicating the usefulness of constrained multi-agent optimization.

\subsection{Effectiveness of Cost Margins}
\label{sec:entropy}

\begin{figure*}[!ht]
    \centering
    \includegraphics[width=\textwidth]{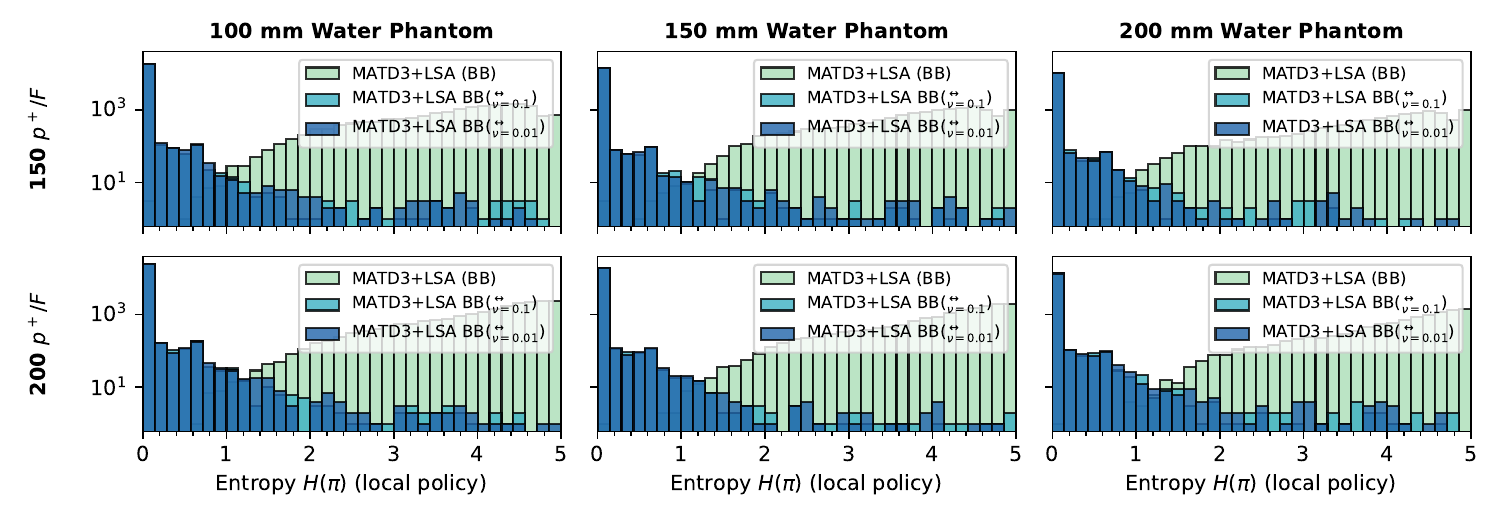}
    \caption{Distributions of the uncertainties in local policy predictions, measured as the predictive entropy for various water phantoms and particle densities. Techniques with enforced cost margins demonstrate significantly reduced uncertainties.}
    \label{fig:policy_entropy}
\end{figure*}

\begin{figure*}[t!]
    \centering
    \includegraphics[width=0.49\linewidth]{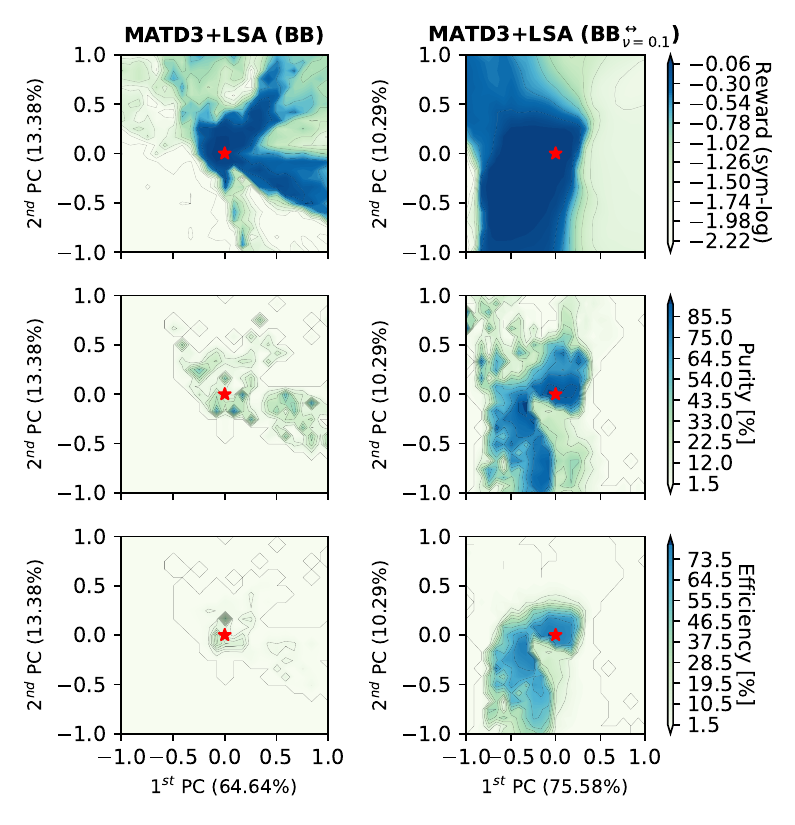}
    \includegraphics[width=0.49\linewidth]{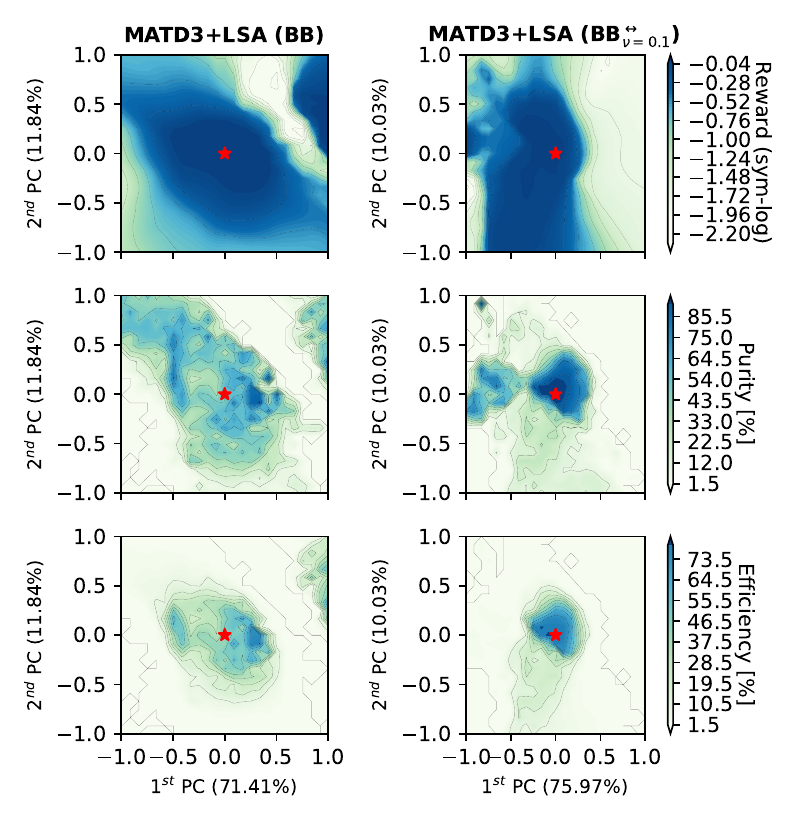}
    \caption{Two-dimensional reward and performance surfaces of multi-agent framework with (MATD3+LSA(BB$^\leftrightarrow_{\nu=0.1}$)) and without cost margins (MATD3+LSA(BB)) generated along the first two principal directions, calculated over the intermediate training checkpoints. Marked with {\color{red}$\bigstar$} are the trained network parameters.}
    \label{fig:reward_surface_margins}
\end{figure*}

\begin{figure*}[t!]
    \centering
    \includegraphics[width=0.49\linewidth]{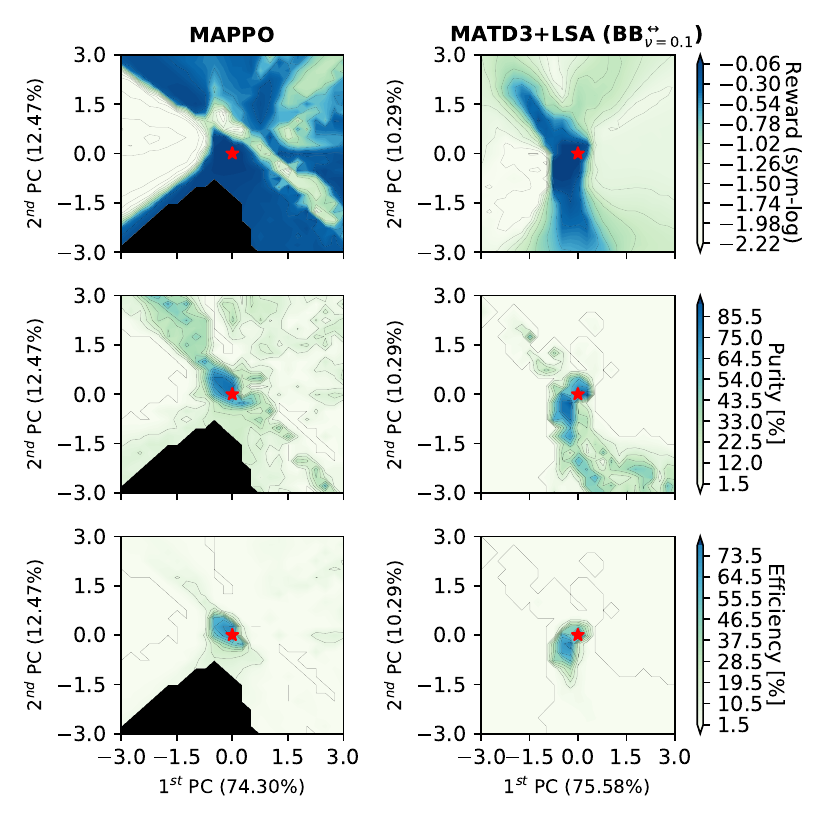}
    \includegraphics[width=0.49\linewidth]{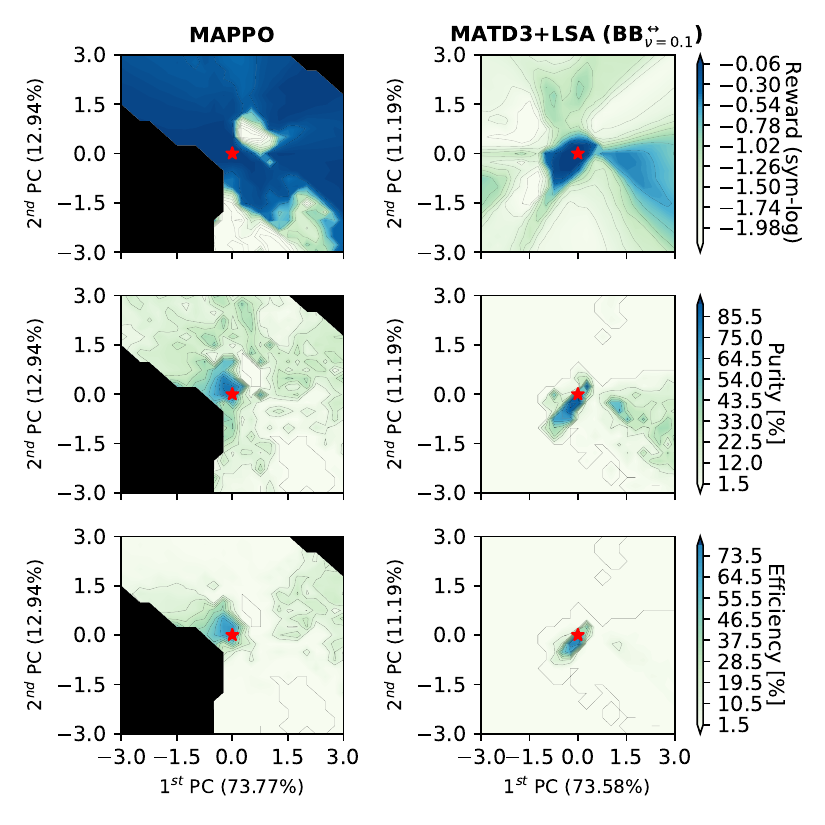}
    \caption{Two-dimensional reward and performance (purity and efficiency) surfaces of multi-agent framework with (MATD3+LSA(BB$^\leftrightarrow_{\nu=0.1}$)) and without policy constraints (MAPPO) generated along the first two principal components, calculated over the intermediate training checkpoints. Marked with {\color{red}$\bigstar$} are the trained network parameters.}
    \label{fig:reward_surface_deg}
\end{figure*}

We verify the effectiveness of the enforced cost margins, described in Section~\ref{sec:architecture}, by analyzing the predictive entropy of the learned policies. Figure~\ref{fig:policy_entropy} shows the distribution of the agents' local policies estimated over all decisions generated over a subset of the first five environments in the dataset for multiple particle density and phantom configurations. We find that local agent policies trained without enforced cost margins show the highest predictive uncertainties (Avg. entropy $\overline{H}(\mu) = 4.099\pm0.221$), indicating only minimal separation from the decision boundaries. For both parameter values of $\nu$, weighing the cost-margin gradient, the long tail of the distribution is reduced significantly, lowering the average entropy by multiple orders of magnitude ($\overline{H}(\mu_{\nu=0.01}) = 0.241\pm0.002$ and $\overline{H}(\mu_{\nu=0.1}) = 0.022\pm 0.003$). We find, similar to the results in Table~\ref{tab:results_all}, that the reduction in uncertainty is robust to the exact choice of $\nu$, showing only marginal different values that are likely due to random mechanisms during training. 

\subsection{Analysis of Policy Constraints and Cost Margins}
\label{sec:surfaces}

The following section presents analyses of reward surfaces for different agents, together with their corresponding surfaces of reconstruction performance. By comparing the reward surfaces with the track reconstruction performance, we aim to  compare and highlight discrepancies in optimization and generalization. Understanding these differences allows us to explain why certain agents, despite achieving similar rewards during training, exhibit vastly different outcomes in terms of reconstruction quality, highlighting the importance of policy constraints as well as cost margins. We generate all surfaces, based on the technique described in~\cite{Li2018c, Sullivan2022}, as two-dimensional slices through the high dimensional landscapes along two directions defined by $\boldsymbol{\nu}$ and $\boldsymbol{\eta}$ according to

\begin{equation}
     f(\alpha, \beta) = \mathcal{L}(\boldsymbol{\theta}^* + \alpha\boldsymbol{\nu} + \beta\boldsymbol{\eta)}.
\end{equation}

We parameterize $\boldsymbol{\nu}$ and $\boldsymbol{\eta}$ as the first two principal components over the entirety of saved training checkpoints (updated every three training iterations). All figures are generated for the 100~$p^+/F$, \SI{100}{mm} phantom dataset with a resolution of, 25$\times$25 uniformly sampled parameter configurations in a region of $[-1, 1] \times  [-1, 1]$ for cost margins and $[-3, 3] \times  [-3, 3]$ for constrained and unconstrained policies. In the latter we experienced multiple configurations where the policy showed numerical issues, resulting in the prediction of \texttt{nan} values, marked in black. \\

\paragraph{Cost Margins} Analyzing the characteristic structure of reward and performance surfaces in~Figure~\ref{fig:reward_surface_margins}, we confirm the initial finding in Section~\ref{sec:results}, that enforcing cost margins with the additional gradient term in Section~\ref{sec:architecture} significantly improves both optimization and generalization. Although the reward surfaces for policies with and without cost margins exhibit a similar shape, we observe a substantial difference in the surfaces for purity and efficiency. We find that the agents with cost margins converge to regions, characterized by wider and stable maxima, suggesting a better generalization performance and a reduced complexity during training. \\

\paragraph{Policy Constraints} Figure~\ref{fig:reward_surface_deg} visualizes the differences in learning abilities for the unconstrained MAPPO and constrained MATD3+LSA architecture with cost margins. Here, we find similarly to Figure~\ref{fig:reward_surface_margins} good agreement of the reward surfaces, while the unconstrained policy shows wider regions of high reward. However, the received reward correlates only moderately with the reconstruction performance, demonstrating a strong degeneracy of the reward surface introduced by the larger combinatorial space caused by unconstrained assignments. Due to misaligned reward signals, the unconstrained agents demonstrate a significant decline in performance, governed by random effects during training (see~Table~\ref{tab:results_all}), indicating the necessity of policy constraints.

\subsection{Functional Similarities and Prediction Instabilities}
\label{sec:instability} 

While both, post-training centralized single-agent (PPO+LSA) and per design centralized multi-agent policies (MATD3+LSA), achieve comparable reconstruction performances, a remaining key question is, whether the two approaches learn similar reconstruction policies and how stable the optimization and final learned policies are, e.g., across random initializations. To quantify potential prediction instabilities~\cite{Mahdi2016, Klabunde2023}, we closely follow the techniques in~\cite{Mahdi2016}~and~\cite{Klabunde2023b}, where the amount of disagreement between two predictors $f_1$ and $f_2$ is quantified as the average fractions of classification errors, defined as 

\begin{equation}
    d = \underset{x, f_{1,2}}{\mathbb{E}} \left[\mathbbm{1} \left\{ \arg \max f_1(x) \neq  \arg \max f_2(x) \right\}\right].
\end{equation}

\cite{Klabunde2023b} proposes an additional extension (min-max normalized disagreement), mapping the raw disagreement rates to a value range of [0, 1] providing better interpretability over the initial approach in~\cite{Mahdi2016}. Following this definition, $d_{\text{norm}}(f_1, f_2)$ is calculated according to

\begin{equation}
        d_{\text{norm}}(f_1, f_2) =  \frac{d(f_1, f_2) - \min d(f_1, f_2)}{\max d(f_1, f_2) - \min d(f_1, f_2)},
\end{equation}

with $\min d(f_1, f_2) = \vert  q_{Err}(f_1) - q_{Err}(f_2) \vert$ and $\max d(f_1, f_2) = \min\left(q_{Err}(f_1) + q_{Err}(f_2), 1\right)$, where $q_{Err}$ is the error rate of a model. However, due to the sequential nature of reinforcement learning, the presented concept of quantifying prediction instabilities not directly applicable, as different predictions lead to changing track candidates. We thus calculate the prediction instability for all manually constructed correctly assigned states, avoiding the propagation of errors throughout the whole detector.

\begin{figure}
    \centering
    \includegraphics[width=\linewidth]{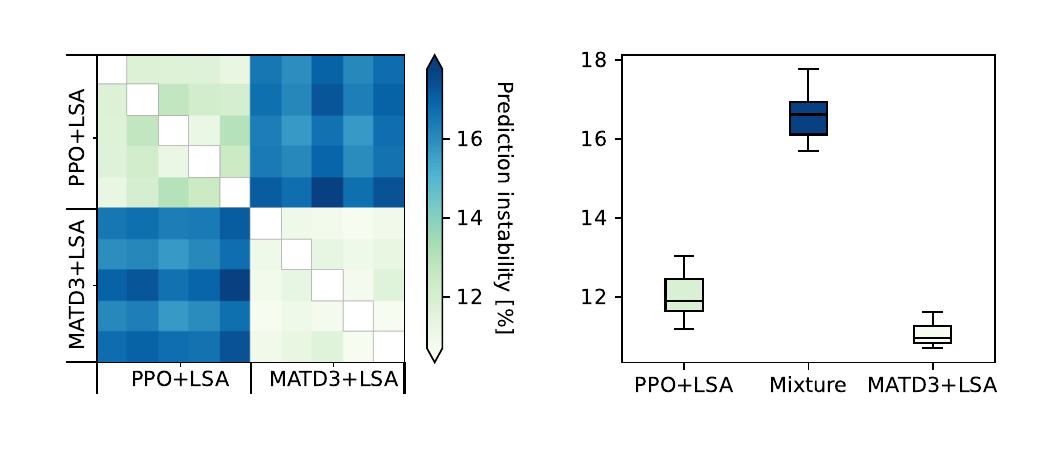}
    \caption{Prediction instabilities of trained reconstruction policies generated for different combinations of optimization algorithm and random initializations.}
    \label{fig:similarities}
\end{figure}

\begin{figure}
    \centering
    \includegraphics[width=\linewidth]{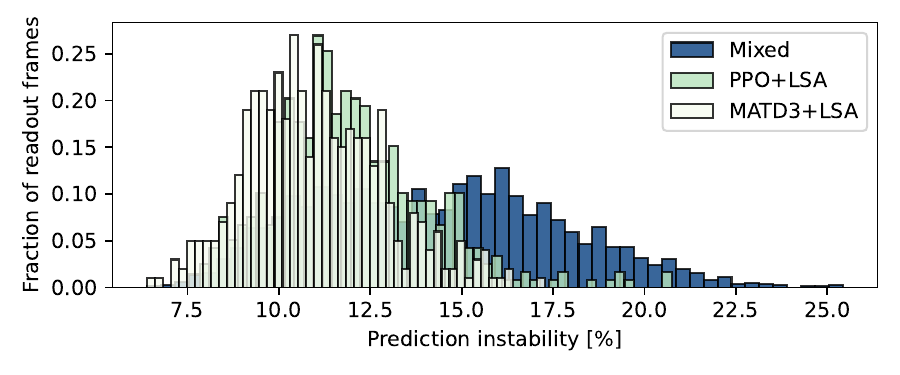}
    \caption{Distributions of prediction instabilities on a readout frame level generated for all combinations presented in Figure~\ref{fig:similarities}.}
    \label{fig:similarities_dist}
\end{figure}

Figure~\ref{fig:similarities} shows both the full correlation-like instability matrix for all combinations of trained agents across agent type and random initializations, as well as the grouped distribution of values. We find that PPO and MATD3+LSA show pronounced differences in training behavior, resulting in substantial prediction instabilities, with a median of approximately 16.5\%. Across different random initializations of the same agent type, we find that the instabilities are reduced. Here, the by design centralized agent demonstrates lower instabilities with an average difference of $0.98$ pp (p-value: 0.01).  While this difference is minor at the presented state, we argue that by the flexibility introduced by team rewards, this effect can be further enhanced. Further we find that while the average prediction instability is considerably low, outliers on a frame-by-frame level, in the form of a long tail of the otherwise Gaussian distribution (see Figure~\ref{fig:similarities_dist}), demonstrate more pronounced instabilities for complex readout frames, posing additional risk for the reconstruction of complex readout frames. Here, we find that our multi-agent approach is able to reduce the number of outliers more effectively compared to the single-agent approach.  

\section{Conclusion}

In this paper, we introduce multiple extensions to an existing single-agent reinforcement learning scheme for charged particle tracking, enabling the joint reconstruction of particle tracks in a multi-agent setting with additional (optional) assignment constraints. We realize the assignment constraints by an implicit, centralized safety layer, projecting the local unsafe actions onto global safe actions. Demonstrating the strong empirical performance of our approach on simulated data for a detector prototype designed for proton, computed tomography, we show that constrained optimization provides an immense advantage over its unconstrained MARL counterpart,  as unconstrained approaches fail to converge consistently to good solutions, due to (1) the high degeneracy of solutions that maximize the team reward signal, while producing a significant amount of incorrect tracks and (2) the increased complexity of spacial credit assignment, most likely introduced by the significantly larger action space of the unconstrained problem. While we were able to achieve similar performance for a post-hoc centralized agent that was trained in a single agent manner, we find that learning particle tracking with constraints reduces the predictive instability, across random initializations. Additionally, using MARL during training provides more flexibility than RL and enables the design of more sophisticated reward functions utilizing information that can be only obtained collaboratively for an aggregate over multiple particle tracks in a readout frame. With the results presented, we aim to extend this work to a generalized and adaptive particle tracking framework that can learn policies for different particle/tracking detectors with additional components, e.g., magnetic fields and is also able to adapt to dynamic changes introduced by, e.g., aging of the detector components.

\section*{Acknowledgements}

This work was supported by the German federal state Rhineland-Palatinate (Forschungskolleg SIVERT) and by the Research Council of Norway (Norges forskningsråd) and the University of Bergen, grant number 250858. TK and NRG gratefully acknowledge the funding of the German National High-Performance Computing (NHR) association for the Center NHR South-West. JK is supported by the Alexander-von-Humboldt-Stiftung. The simulations and computations were executed on the high performance cluster "Elwetritsch" at the University of Kaiserslautern-Landau (RPTU), which is part of the "Alliance of High Performance Computing Rhineland-Palatinate" (AHRP). We kindly acknowledge the support of the regional university computing center (RHRK). The ALPIDE chip was developed by the ALICE collaboration at CERN.

\section*{Members of the Bergen pCT Collaboration}
{
\scriptsize\noindent
Max Aehle\textsuperscript{a},
Johan Alme\textsuperscript{b},
Gergely Gábor Barnaföldi\textsuperscript{c},
Tea Bodova\textsuperscript{b},
Vyacheslav Borshchov\textsuperscript{d},
Anthony van den Brink\textsuperscript{b},
Mamdouh Chaar\textsuperscript{b},
Viljar Eikeland\textsuperscript{b},
Gregory Feofilov\textsuperscript{f},
Christoph Garth\textsuperscript{g},
Nicolas R.\ Gauger\textsuperscript{a},
Georgi Genov\textsuperscript{b},
Ola Grøttvik\textsuperscript{b},
Håvard Helstrup\textsuperscript{h},
Sergey Igolkin\textsuperscript{f},
Ralf Keidel\textsuperscript{a,i},
Chinorat Kobdaj\textsuperscript{j},
Tobias Kortus\textsuperscript{a},
Viktor Leonhardt\textsuperscript{g},
Shruti Mehendale\textsuperscript{b},
Raju Ningappa Mulawade\textsuperscript{i},
Odd Harald Odland\textsuperscript{k, b},
George O'Neill\textsuperscript{b},
Gábor Papp\textsuperscript{l},
Thomas Peitzmann\textsuperscript{e},
Helge Egil Seime Pettersen\textsuperscript{k},
Pierluigi Piersimoni\textsuperscript{b,m},
Maksym Protsenko\textsuperscript{d},
Max Rauch\textsuperscript{b},
Attiq Ur Rehman\textsuperscript{b},
Matthias Richter\textsuperscript{n},
Dieter Röhrich\textsuperscript{b},
Joshua Santana\textsuperscript{i},
Alexander Schilling\textsuperscript{a},
Joao Seco\textsuperscript{o, p},
Arnon Songmoolnak\textsuperscript{b, j},
Ákos Sudár\textsuperscript{c, q},
Jarle Rambo Sølie\textsuperscript{r},
Ganesh Tambave\textsuperscript{s},
Ihor Tymchuk\textsuperscript{d},
Kjetil Ullaland\textsuperscript{b},
Monika Varga-Kofarago\textsuperscript{c},
Boris Wagner\textsuperscript{b},
RenZheng Xiao\textsuperscript{b, v},
Shiming Yang\textsuperscript{b},
Hiroki Yokoyama\textsuperscript{e},

\smallskip\noindent
a) Chair for Scientific Computing, University of Kaiserslautern-Landau (RPTU), 67663 Kaiserslautern, Germany
b) Department of Physics and Technology, University of Bergen, 5007 Bergen, Norway;
c) Wigner Research Centre for Physics, Budapest, Hungary;
d) Research and Production Enterprise ''LTU'' (RPELTU), Kharkiv, Ukraine;
e) Institute for Subatomic Physics, Utrecht University/Nikhef, Utrecht, Netherlands;
f) St.\ Petersburg University, St.\ Petersburg, Russia;
g) Scientific Visualization Lab, University of Kaiserslautern-Landau (RPTU), 67663 Kaiserslautern, Germany;
h) Department of Computer Science, Electrical Engineering and Mathematical Sciences, Western Norway University of Applied Sciences, 5020 Bergen, Norway;
i) Center for Technology and Transfer (ZTT), University of Applied Sciences Worms, Worms, Germany;
j) Institute of Science, Suranaree University of Technology, Nakhon Ratchasima, Thailand;
k) Department of Oncology and Medical Physics, Haukeland University Hospital, 5021 Bergen, Norway;
l) Institute for Physics, Eötvös Loránd University, 1/A Pázmány P. Sétány, H-1117 Budapest, Hungary;
m) UniCamillus -- Saint Camillus International University of Health Sciences, Rome, Italy;
n) Department of Physics, University of Oslo, 0371 Oslo, Norway;
o) Department of Biomedical Physics in Radiation Oncology, DKFZ—German Cancer Research Center, Heidelberg, Germany;
p) Department of Physics and Astronomy, Heidelberg University, Heidelberg, Germany;
q) Budapest University of Technology and Economics, Budapest, Hungary;
r) Department of Diagnostic Physics, Division of Radiology and Nuclear Medicine, Oslo University Hospital, Oslo, Norway;
s) Center for Medical and Radiation Physics (CMRP), National Institute of Science Education and Research (NISER), Bhubaneswar, India;
t) Biophysics, GSI Helmholtz Center for Heavy Ion Research GmbH, Darmstadt, Germany;
u) Department of Medical Physics and Biomedical Engineering, University College London, London, UK;
v) College of Mechanical \& Power Engineering, China Three Gorges University, Yichang, People's Republic of China
\par
}

\bibliographystyle{IEEEtran}
\bibliography{IEEEabrv,main.bib}

\vskip -2\baselineskip plus -1fil

\begin{IEEEbiography}[{\includegraphics[width=1in,height=1.25in,clip,keepaspectratio]{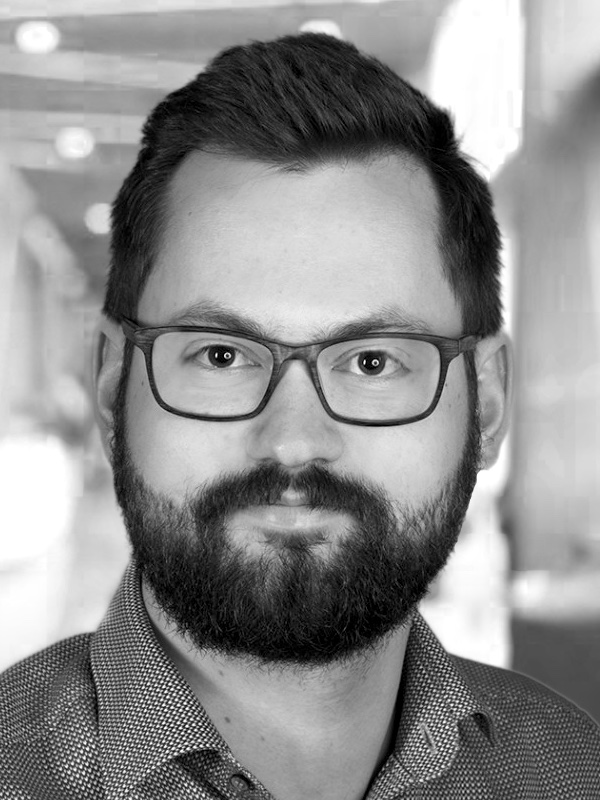}}]{Tobias Kortus} received the B.Sc. degree in Medical Engineering from University of Applied Sciences Furtwangen, University Campus Tuttlingen, in 2019 and the M.Sc. degree in Applied Computer Science from University of Applied Sciences Esslingen in 2021. He is currently working towards his PhD at the University of Kaiserslautern-Landau. His research interests include machine learning and reinforcement learning, with focus on applications in high energy and medical physics. 
\end{IEEEbiography}

\vskip -2\baselineskip plus -1fil

\begin{IEEEbiography}[{\includegraphics[width=1in,height=1.25in,clip,keepaspectratio]{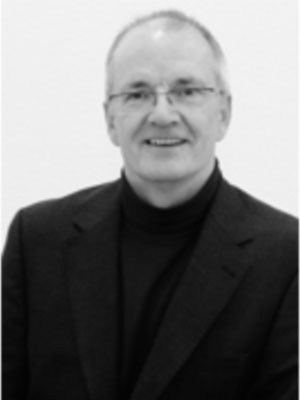}}]{Ralf Keidel} is Senior Professor at the University of Applied Sciences Worms and Principal Investigator of the SIVERT research training group dealing with the algorithmic part of the proton Computed Tomography (pCT) project of the Bergen pCT Collaboration. He is member of the ALICE collaboration board and the Inter-experimental Machine Learning Working Group at CERN, Geneva. His research interests are pCT, machine learning and optimization techniques.
\end{IEEEbiography}

\vskip -2\baselineskip plus -1fil

%\newpage
\begin{IEEEbiography}[{\includegraphics[width=1in,height=1.25in,clip,keepaspectratio]{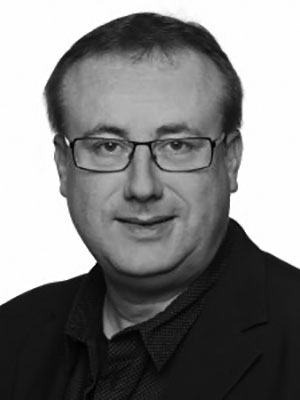}}]{Nicolas R. Gauger}
Nicolas R. Gauger is Full Professor and Chairholder for Scientific Computing and Director of the Computing Center (RHRK) at University of Kaiserslautern-Landau as well as Principal Investigator of the SIVERT research training group dealing with the algorithmic part of the proton Computed Tomography (pCT) project of the Bergen pCT Collaboration. His research interests are numerical optimization, high-performance computing, machine learning and pCT amongst other fields of application. 
\end{IEEEbiography}

\vskip -2\baselineskip plus -1fil

\begin{IEEEbiography}[{\includegraphics[width=1in,height=1.25in,clip,keepaspectratio]{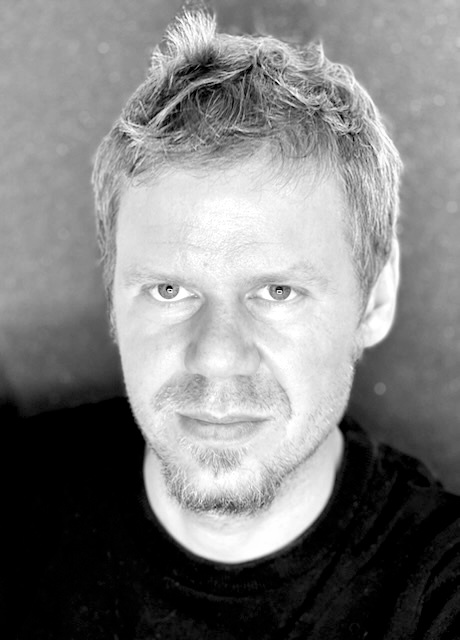}}]{Jan Kieseler}

Jan Kieseler is a junior group leader at KIT, focusing on top quark physics and upgrades in the CMS experiment, and on developing particle reconstruction algorithms, including object identification techniques used in CMS, Belle II, and FCC collaborations. In addition to several leadership roles in  top quark physics and upgrades, he co-founded the CMS ML4RECO initiative and is a founding member of the MODE collaboration.
\end{IEEEbiography}

\vskip -2\baselineskip plus -1fil
\vskip -2\baselineskip plus -1fil
\vfill

\end{document}